\documentstyle[prd,aps,floats,epsfig]{revtex}
 
\begin{document} 

\def\be{\begin{equation}}
\def\ee{\end{equation}}
\def\ba{\begin{eqnarray}}
\def\ea{\end{eqnarray}}
 
\author{Carlo R. Contaldi} 
\address{Theoretical Physics, The Blackett Laboratory, 
Imperial College, Prince Consort Road, London, SW7 2BZ, U.K.\\ 
} 
\twocolumn[\hsize\textwidth\columnwidth\hsize\csname@twocolumnfalse\endcsname 
 
\title{Cosmic Strings in the age of Boomerang} 
 
\maketitle 
\begin{abstract} 
We show how two simple modifications to the standard cosmic string
scenario for structure formation compare to the recently released
Boomerang data set. Namely we consider pure string closed models and
mixed models where both inflation and strings are responsible for the
perturbations. In the closed models we find that pure string models
would require a universe with roughly $\Omega_M=0.8$,
$\Omega_{\Lambda}=1.6$ to agree with the peak position revealed by the
data and in agreement with the SNIa data.
In the hybrid scenario with local strings we find that
we require even more tilt and baryon content ($n_s\sim 0.8$,
$\Omega_b=0.08$, $H_0=70$) to match the data
than with pure inflation models. The case with global strings fares
better with a standard period of $\Lambda$CDM inflation and a $\sim30\%$
contribution from strings being in good agreement with the data.
\end{abstract} 
\date{\today} 
\pacs{PACS Numbers: 98.80.Cq, 98.80.-k, 95.30.Sf} 
] 
 

The recently published Boomerang data \cite{ldb} has mapped the peak
in the CMB power spectrum to new accuracies. The peak position and
shape heavily support the idea that the perturbations at angular
scales around $\ell \sim 200$ are dominated by acoustic fluctuations
of the photon-baryon fluid at last scattering. In fact the observation
of a peak in such agreement to that predicted for the tight coupling
adiabatic scenario will probably come to be recognized as one of the major 
achievements of modern theoretical and observational cosmology.

In an inflationary context, the models most favoured
by the Boomerang data, for a flat universe, seem to be $\Lambda$CDM models with
a high baryon content
or considerable spectral tilt $n_s$ in the primordial power spectrum
\cite{ldb1}. The best fit models strongly depend on the priors imposed
from other cosmological observations such as measurements of $h$ and
nucleosynthesis limits on $\Omega_bh^2$. Imposing strong constraints
on these parameters seems to indicate slightly closed models with
$\Omega_{\Lambda}\sim 0.7$,
$\Omega_b\sim 0.05$, $\Omega_{C}\sim0.3$ $H_0\sim70$ and $n_s\sim0.85$. The
indication that the spectral index might not be very close to unity
may rule out some of the simplest and most appealing models of inflation.

The presence of such a well defined peak at $\ell \sim 200$ heavily
disfavours the standard topological defect scenarios (see
e.g. \cite{usg,ruth,turok}) which have been the the classic rivals to
inflation theories in the structure formation debate over the past
decade. Passive models of structure formation such as those where the
perturbations are set up by inflation produce acoustic peaks in the
CMB power spectrum because the equations of motion are homogeneous and
the perturbations linear so that the coherence of the initial
perturbations is preserved. In active defect theories such as cosmic strings
the coherence is lost because the perturbations are being sourced
continuously by the non-linearly evolving stress-energy of the string
network. The acoustic peaks are therefore not ususally
present in defect theories unless some extreme coherence conditions
are imposed on the defect evolution \cite{turok1}.

Theories for global defects \cite{ruth,turok} predict a wide peak for the
CMB power spectrum due to the relatively 
large contribution to the total $C_{\ell}s$ from the vector and tensor
modes which tends to smear out the rise in the spectrum relative to
the Sachs-Wolfe plateau. On the other hand theories for local cosmic
strings seem to predict a more pronounced peak due to the relative
suppression of the tensor and vector modes but its position is
consistently shifted to higher scales with $\ell_{peak} \sim 450-600$
\cite{usg,andy} in obvious disagreement with the Boomerang data. 

\begin{figure}
\centerline{\psfig{file=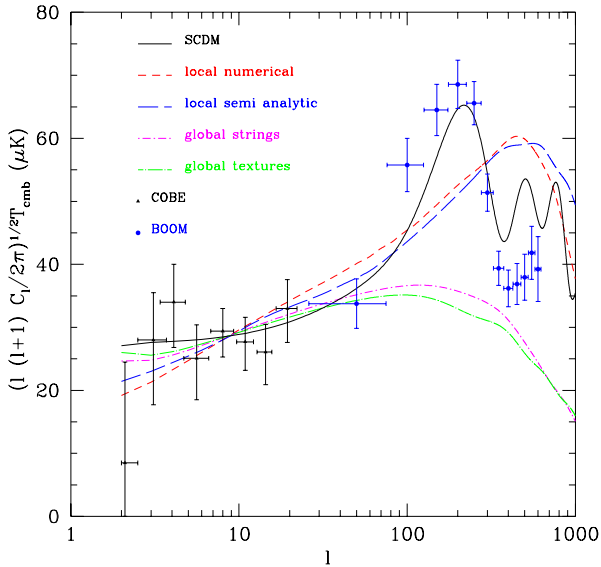,width=8 cm,angle=0}}
\caption{The CMB power spectra for local strings and global strings and
texture models compared with the Boomerang and  COBE 4-year data
points.} 
\label{stand}
\end{figure}

In Fig~\ref{stand} we show various spectra from defect theories
plotted with the COBE and Boomerang data sets. The `local numerical' curve
indicates a typical spectrum for local cosmic strings obtained via
numerical simulations of Nambu-Goto strings \cite{usg}. The spectrum is
obtained by sourcing the perturbations with a complete set of
scaling correlation functions for the components of the
network's stress-energy tensor. The scaling correlators are measured
directly from simulations. The `global
strings' and `global texture' curves show examples of a spectrum
obtained using global 
defect simulations \cite{turok}.  The `semi-analytic' curve shows a
comparable spectrum from \cite{andy1} where a semi-analytic model is
employed to produce histories for the sourcing terms. A  
standard CDM inflationary spectrum is included for comparison. 

One of the most interesting aspects of the data that has already
generated much interest \cite{white1,lesg,peebles} is the low power at scales
$\ell > 350$ and the absence of a pronounced secondary peak. As
pointed out in the initial reactions to the data, this might indicate
a tilted spectrum or a higher than expected baryon content. It has also
been suggested \cite{peebles} 
that topological defects might be invloved in explaining this {\it
apparent} discrepancy from the standard $\Lambda$CDM scenario by
delaying the onset of recombination. 

The aim of this {\it letter} is to carry out an initial survey of
possible non-standard cosmic string scenarios and establish whether
they are still viable or indeed whether they might help to explain this effect.


An initial, admittedly, crude attempt to bring the local cosmic string
spectra back into line with the data is to assume a closed
universe. In analogy with the peak position in inflation models a
closed geometry would shift scales to lower $\ell$s. The angular size
of the peaks in both inflation and defect theories are relatively
insensititve to the cosmological parameters \cite{joao,bond} except
for the total energy denisity $\Omega$. In inflation $\ell_{peak}$
depends on both the size of the sound horizon at decoupling $d_s(z_*)$
which limits the largest wavelength which has had time to start
oscillating and the angular diameter distance to the last scattering
surface $d_A(z_*)$ which tells us how the angular size of an object is
affected by the geometry of the space on the line of sight. For
$\Omega$ close to unity and high redshifts these are insensitive
to the individual energy density components and the peak position
scales simply as 
\be
	\ell_{peak} \sim \frac{d_A(z_*)}{d_s(z_*)} \approx \frac{200}{\Omega^{1/2}}.
\ee

In the case of cosmic strings the peak position in an $\Omega = 1$
universe depends on the detailed structure of the network
stress-energy forcing the perturbations \cite{usg,joao,shel,aust} and is in
general at smaller angular scales than the scale of the sound
horizon. As a first approximation to a spectrum for a closed universe
we can therefore 
shift angular scales according to $d_A(z_*)$. Since we expect large
($\sim1$)
deviations from unity for $\Omega$,  $d_A(z_*)$ will now depend
explicitly on the different matter components and their relative
contributions and requires numerical integration. 
 
\begin{figure}
\centerline{\psfig{file=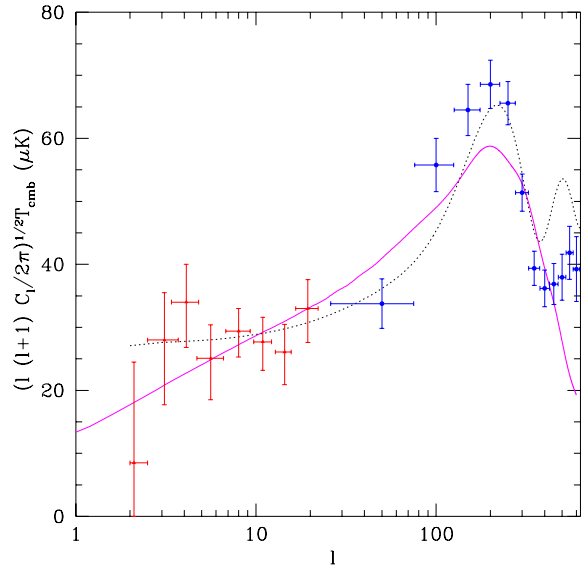,width=8 cm,angle=0}}
\caption{An $\Omega_{\Lambda}=0.7$, $\Omega_M=0.3$ spectrum shifted to
illustrate the effect of overcritical density(in 
this case an $\Omega_{\Lambda}=1.6$,
$\Omega_M=0.8$ and $H_0=70$). Notice that in practice the spectrum
would be flatter at large scales and a change in $\Omega$ would also
significantly change the damping at small scales, possibly broadening
the peak somewhat although it is hard to see how the spectrum might
reproduce the structure in the data at $\ell > 350$} 
\label{close}
\end{figure}

In Fig.~\ref{close} we show how the projection would shift the peak to
the appropriate scales. The spectrum being used is that obtained for
the evolution of a cosmic string network in a flat $\Lambda$CDM universe
using a modified Einstein-Boltzmann code \cite{usg,joao} with
$\Omega_{\Lambda}=0.7$, $\Omega_M=0.3$ and $H_0=70$. The shift shown
in Fig.~\ref{close} is obtained by
increasing the $\Lambda$ contribution to
$\Omega_{\Lambda}=1.6$ and the total matter contribution to
$\Omega_M=0.8$. The shift is degenerate with respect to
$\Omega_{\Lambda}$ and  $\Omega_M$. The values shown is
the most `conservative' model which is within the $68\%$ contour of
the SNIa \cite{perl} $\Omega_{\Lambda}$, $\Omega_M$ confidence plot
and gives an accelerating expansion at the present epoch.

As usual when dealing with a closed universe one
has to be careful about 
the age of the universe which is too small for large $\Omega_M$ values
but including a relatively large contribution from the cosmological
constant gets around the problem with the above closed model giving a
total age $T_0\sim15$ Gyrs and $\Lambda$ domination occuring at
redshifts $z \sim 0.4$. Other traditional constraints against closed
models such as peculiar velocities or gravitational lensing limits
(see e.g. \cite{white}) are
usually derived in an adiabatic 
CDM context and hence do not relate straightforwardly to a closed
defect model and require a more detailed treatment.

This simple example highlights the fact that there is an unexplored
region in the ($\Omega_{\Lambda}$, $\Omega_M$)  parameter space given
by $\Omega > 1$, $\Omega_{\Lambda} > 1$ and 
$\Omega_M < 1$  which is very relevant for topological defects given
the Boomerang results and which should be investigated
further. Naturally Fig.~\ref{close} serves only as an 
illustration of the effect as the details of the spectrum can only be
obtained by a numerical treatment of actively sourced perturbations in
closed universes which is currently unavailable. In particular the
ISW scales would not be affected to such a large degree greatly
improving the tilt  at COBE scales. Also the high $\ell$ tail of the
spectrum will change considerably as damping effects are not expected
to scale back so simply, intuitively we'd expect to see a
slightly broader peak. Even so it is hard to see how the
spectrum might reproduce the structure already observed in the data at
$\ell > 350$.


Another approach which has generated interest recently is to have a
model where both inflation and defects are 
responsible for seeding and sourcing cosmological
perturbations. This can occur in hybrid inflation models (see e.g. \cite{linde,rich}) where one field is responsible for the potential energy
dominated era and the other provides a non-zero VEV, after a suitable
symmetry breaking phase, for the production of topological
defects. These models were initially seen to suffer from an increased 
fine tuning problem due to the requirment that the defects were formed at a
sufficiently late stage or at the end of inflation. Their
extensions to supersymmetric models of inflation and in particular to
supergravity theories however do not suffer as badly from this
probelm as the parameters in the theory seem to have more natural  
justifications in superstring theory \cite{lyth}.

An example of such a model is $D$-term inflation \cite{rach,us1}
where a symmetry breaking  
phase occurs after inflation in one of the flat directions in the
potential. In particular, supergravity theories tend to favour $D$-term
inflation where local cosmic strings are produced when an extra
gauged U(1) symmetry is broken. The energy per-unit length $\mu$ of the
strings produced in D-term inflation is related to the
Fayet-Iliopoulos term $\xi$ as $\mu = 2\pi\xi$. This parameter is essentially
added in by hand in anomaly free models in order to obtain symmetry
breaking but it may be related to the gauge 
coupling $g_{str}$ of weakly coupled string theory when an anomalous $U(1)$
symmetry is present. The strings are expected to contribute
significantly to the 
CMB spectrum \cite{rach}.

As pointed out in \cite{us1,rich,rich1}, if we assume that the two
perturbations are uncorrelated and that the strings are formed
sufficiently late during 
inflation so as not to be diluted significantly, obtaining the total CMB
power spectra is trivial with, 
\be
	C_{\ell}^{tot} = \alpha C_{\ell}^{inf} + (1-\alpha)C_{\ell}^{str}
\ee 
where $\alpha$ is the parameter giving the relative contribution from
inflation versus strings and the two spectra are normalized to
COBE. The parameter $\alpha$ can vary between $0$ and $1$ and is related to the number of e-foldings
and slow roll parameters $\epsilon$ and $\eta$ \cite{rach}. It may
therefore help to  
constrain the form of the potential. In \cite{rach} it was shown how a
particular implementation 
of D-term inflation gives $\alpha\sim 0.25$ for $N=60$ e-foldings. The
assumption that the two effects are  
indeed uncorrelated in this simple scenario is well motivated since
cosmic string networks 
have very short coherence times and lose memory of their initial conditions
very quickly. 

\begin{figure}
\centerline{\psfig{file=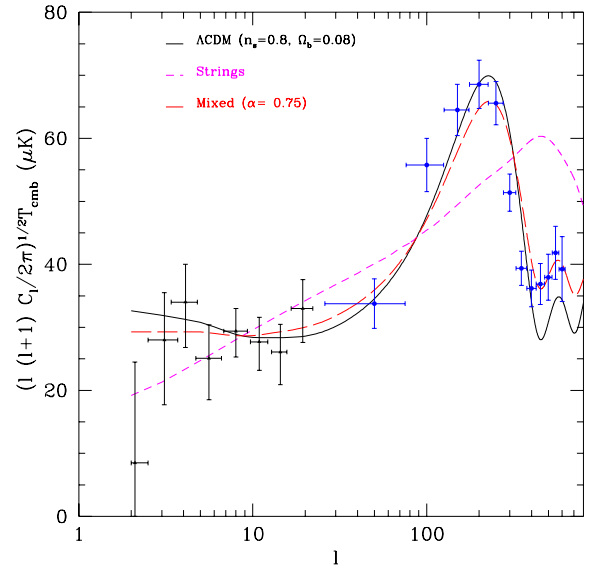,width=8 cm,angle=0}}
\caption{The hybrid scenario with local cosmic strings. Both the
string and inflation spectra are those of an $H_0=70$,
$\Omega_{\Lambda}=0.7$, $\Omega_b=0.8$ model with a primordial power
spectrum tilt of $n_s=0.8$ for inflation. The extreme values are
required to bring the power at $\ell > 350$ down to acceptable levels
when mixing contributions from both spectra.} 
\label{local}
\end{figure}

An interesting development is in the case of local strings, the new
Boomerang data restricts the 
possible mixing scenarios quite strongly. 
Fig.~\ref{local} shows a very good fit to the data for a mixed model
but this is quite deceptive as it requires a heavily tilted inflation
spectrum ($n_s=0.8$) and a  high baryon content ($\Omega_b=0.08$) to
suppress the power even further below the 
Boomerang data at high $\ell$s. The reason for this is simple, the local string
spectrum peaks at scales $\ell > 350$ so that even modest amounts of
string contributions ($\alpha \sim 0.8$) will only make things worse
for the low power problem that the data is apparently implying. In
effect one has to tweak the inflation spectrum even further than in a pure
inflation scenario to make the mixed scenario work. 
  
\begin{figure}
\centerline{\psfig{file=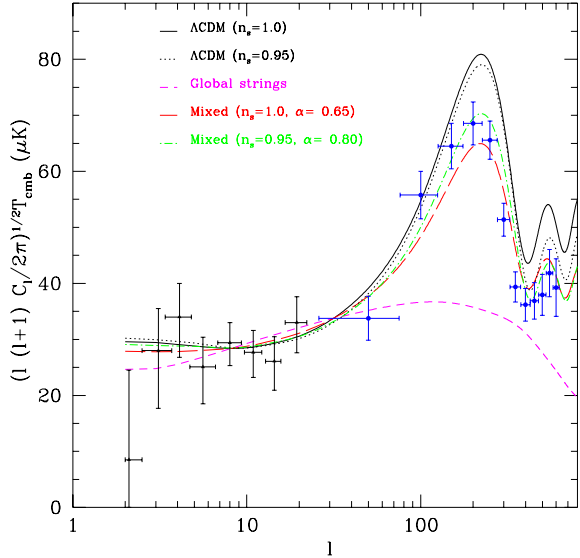,width=8 cm,angle=0}}
\caption{A hybrid scenario with global strings. The two inflation spectra
are for standard $\Lambda$CDM models ($\Omega_{\Lambda}=0.7$,
$\Omega_b=0.8$, $\Omega_{C}=0.25$ and $H_0=70$) with $n_s=1.0$ and
$n_s=0.95$. The string spectrum is  an example of a
SCDM global model.} 
\label{global}
\end{figure}

The scenario fares better if global defects are
produced towards the end of the inflationary. Although it is often
stated in the phenomenological 
literature that SUSY inflation results inevitably in local strings
this is not necessarily so \cite{lyth}. Different types and even mixtures
of defects can be produced in such models with the stability  and type
of the defects depending on the exact symmetry breaking. 
In Fig.~\ref{global} we use an example of a global string
spectrum from \cite{turok} (we do not have any $\Lambda$CDM spectra in
the global case but as with the local string case a non-zero
cosmological constant is not expected to alter the shape of the
spectrum drastically). As shown, a considerable contribution from the 
strings to a standard $\Lambda$CDM inflation picture  agrees very
well with the data. We plot the mixed spectrum obtained for two  
$\Lambda$CDM inflation models with $\Omega_{\Lambda}=0.7$,
$\Omega_b=0.05$, $\Omega_{C}=0.25$ but with different spectral indeces
($n_s=1.0$ and $n_s=0.95$). In both cases the drop-off in the global
string spectrum suppresses the power of the secondary peaks with
respect to the first peak which helps reconcile the standard
$\Lambda$CDM model without having to increase the baryon density. This
may prove to be useful if future high $\ell$ data does indeed confirm
a conflict between inflation scenarios and  nucleosynthesis
constraints on $\Omega_b$ or smoothed out (decohered) peaks.

In summary we have shown how two simple modifications to the standard cosmic
string scenario compare to the new CMB data. In the first case we have
seen that closed models may help to bring the position of the peak in
line with the data although we can only speculate how the string model
could reproduce the structure seen in the boomerang data. Hybrid
inflation models,  where both inflation and strings are responsible for
the perturbations, only help in the case with {\it global} defects. Adding
even small amounts of local strings only makes matters worse as the
inflation spectrum has to be `tweaked' to unattractive levels. 

It should be stressed that the hybrid models considered here are
extremely simplistic and the weighted sum mixing should only be
regarded as a useful tool to question the viability of the simplest models. In
fact the evolution of defects produced in supersymmetric theories may
differ quite significantly to the standard models and the defect
scenario might be much more exotic. A likelihood analysis for these mixed
models would be quite premature at this time as we do
not have the necessary tools yet to build a grid of models in the case of
global defects\footnote{During the preparation of this work a separate
group has submitted similar work on the archive \cite{bouch} dealing with hybrid
inflation models involving global strings. The spectrum reported in
\cite{bouch} is essentially that of Fig.~\ref{global}.}. It is encouraging though that as
seen in the local string case the new data may already rule out such
simple models.

We thank Jo\~ao Magueijo, Pedro Ferreira, Richard Battye and Daniele
Steer for useful comments.  Special 
thanks are due to   U-L. Pen, U. Seljak, and N. Turok for the use of
their global defect spectra and to A. Albrecht, R. Battye and J. Robinson
for the use of their semi-analytic results. We acknowledge financial
support from the Beit Fellowship in Science.

\end{document}